\documentclass[pre,aps,twocolumn,superscriptaddress,showpacs]{revtex4-1}
\usepackage{graphicx}
\usepackage{amsmath}

\usepackage{color}

\begin{document}

\title{A universal opportunity model for human mobility}

\author{Er-Jian Liu}
\affiliation{Key Laboratory of Transport Industry of Big Data Application Technologies for Comprehensive Transport, Ministry of Transport, Beijing Jiaotong University, Beijing 100044, China}
\affiliation{Institute of Transportation System Science and Engineering,
Beijing Jiaotong University, Beijing 100044, China}

\author{Xiao-Yong Yan} \email{yanxy@bjtu.edu.cn}
\affiliation{Institute of Transportation System Science and Engineering,
Beijing Jiaotong University, Beijing 100044, China}
\affiliation{Comple$\chi$ Lab, University of Electronic Science and Technology of China, Chengdu 611731, China}

\begin{abstract}

Predicting human mobility between locations has practical applications in transportation science, spatial economics, sociology and many other fields. For more than 100 years, many human mobility prediction models have been proposed, among which the gravity model analogous to Newton's law of gravitation is widely used. Another classical model is the intervening opportunity (IO) model, which indicates that an individual selecting a destination  is related to both the destination's opportunities and the intervening opportunities between the origin and the destination. The IO model established from the perspective of individual selection behavior has recently triggered the  establishment of many new IO class models. Although these IO class models  can achieve accurate prediction at specific spatiotemporal scales,  an IO class model that can describe an individual's destination selection behavior at different spatiotemporal scales is still lacking. Here, we develop a universal opportunity  model that considers two human behavioral tendencies: one is {\color{black} the exploratory tendency}, and the other is {\color{black} the cautious tendency}.  Our model establishes a new framework in IO class models and covers the classical radiation model and opportunity priority selection model. Furthermore, we use various mobility data to demonstrate our model's predictive ability. The results show that our model can better predict human mobility than previous IO class models. {\color{black} Moreover, this model can help us better understand the underlying mechanism of the individual's destination selection behavior in different types of human mobility.}

\end{abstract}


\maketitle

\section{Introduction}

Predicting human, goods and information mobility between locations is an important topic in complex human behavior \cite{Ba19,Ba10}, transportation science \cite{OW11,HU18}, sociology \cite{hu09}, economic geography \cite{RT03} and regional economics \cite{Lm18,Ka00,pa07}, and it also has practical applications in urban planning \cite{ld17,ba08}, population migration \cite{to95}, cargo transportation \cite{ka10}, traffic engineering \cite{he01}, infectious disease epidemiology \cite{hu004,eu004,bal09,wag09} and emergency management \cite{bag11,lu12,ru13}. 
For more than 100 years, researchers have proposed a variety of models for predicting the mobility of people between locations.
The most influential model is the gravity model, which is analogous to  Newton's law of gravitation, i.e., the flow between two places is proportional to their population and decays as the power of their distance.
The gravity model is simple in form and has been successfully used to predict railway freight volume \cite{Z46}, subway passengers \cite{gon12}, highway traffic flow \cite{JWS08}, air travel \cite{gr07}, commuting \cite{VBS06} and population migration \cite{to95}.
Hereafter, researchers derived the gravity model from the perspective of destination selection behavior using the theory of determining utility \cite{N69}, stochastic utility \cite{D75} and game theory \cite{yz19}.
Another classic model that is also established from the perspective of destination selection behavior is  the intervening opportunity (IO) model \cite{S40}.
Different from the gravity model, the IO model takes the total number of opportunities (often proportional to population) between the origin and the destination (named intervening opportunities), instead of the actual distance between the two places, as a key factor in determining human mobility. The concept of intervening opportunities provides a new direction for constructing the human mobility prediction model \cite{BBG17}.
	
Inspired by the IO model, Simini et al. establish a parameter-free human mobility model named the radiation model \cite{SGMB12}. The radiation model assumes that when seeking job offers, the commuter will choose the closest workplace to his/her home, whose benefit is higher than the best offer available in his/her home county, i.e., the benefit of home is higher than the benefits of the intervening opportunities and lower than the benefit of the workplace. The radiation model can better predict the commuting behavior between counties. Some researchers improve the radiation model and propose various commuting prediction models, such as the radiation model with selection \cite{SMN13}, generalized radiation model \cite{KLGQ15}, the flow and jump model \cite{VTN18}, travel cost optimized radiation model \cite{var16} and a cost-based radiation model \cite{REWGT14}. Yan et al. propose a population-weighted opportunities (PWO) model \cite{YZFDW14} by mining human daily travel data from several cities, such as the GPS trajectories from vehicles and call detail records from mobile phones.
The PWO model assumes that the probability of an individual selecting a destination is proportional to the number of opportunities at the destination and inversely proportional to the total population at the locations whose distances to the destination are shorter than or equal to the distance from the individual's origin to the destination, which can better predict intracity trips. Yan et al. further combine the PWO model with the continuous-time random walks model \cite{mon65} to obtain a universal model of individual and population \cite{YWGL17}, which realizes the prediction of intracity and intercity mobility patterns at both the individual and population levels.
{\color{black}Huang et al.  propose a novel human mobility model that can capture real-time human mobility in a sustainable and economical manner, which broadens our view.\cite{HU18}}
Sim et al. establish a deliberate social tie (DST) model \cite{sim15} from the perspective of social interactions. The DST model assumes that an individual seeks out social ties
only with other individuals whose attribute values are higher than the attribute value of the individual and the attribute values of the intervening opportunities. Motivated by the DST model, Liu and Yan propose an opportunity priority selection (OPS) model that assumes that the destination selected
by the individual is the location that presents a higher benefit than the benefit of the origin and the benefits of the intervening opportunities \cite{LY19}. In general, all of the IO class models \cite{SGMB12,SMN13,KLGQ15,VTN18,var16,REWGT14,YZFDW14,YWGL17,sim15,LY19} share two common assumptions: (i)
using an agent to represent all of the individuals; (ii)  when selecting a destination, the agent will compare the benefits of different locations. The difference between these IO class models is that the rules for comparing benefits of different locations are different. Although the radiation  class models \cite{SGMB12,SMN13,KLGQ15,VTN18,var16,REWGT14} can accurately predict commuting behavior and other IO class models \cite{YZFDW14,YWGL17,sim15,LY19} can accurately predict intracity and/or intercity mobility, an IO class model that can simultaneously describe the individual's destination selection behavior at different spatiotemporal scales is still lacking.
	
In this paper, we propose a universal opportunity (UO) model  to characterize an individual's destination selection behavior. The basic idea of the model is that when an individual selects a destination, she/he will comprehensively compare the benefits of the origin, the destination and the intervening opportunities.
Furthermore, we use various mobility data sets to demonstrate the predictive power of our model.
The results show that the model can accurately predict different spatiotemporal scale movements such as intracity trips, intercity travels, intercity freight, commuting, job hunting and migration. Moreover, our model can also cover the classical radiation model and OPS model, presenting a new universal framework for predicting human mobility in different scenarios. 
	
\section{Results}
	
\subsection{Model}
We assume that when an individual chooses a destination, like the radiation model \cite{SGMB12} and the OPS model \cite{LY19}, she/he first evaluates the benefit of the location's opportunities \cite{pan03} where the benefit is randomly chosen from a distribution $p(z)$. After that, the individual comprehensively compares the benefits of the origin, the destination and the intervening opportunities and selects a location as the destination. To characterize the behavior of an individual comprehensive comparison of the benefits of the locations, we use two parameters $\alpha$ and $\beta$.  Parameter $\alpha$ reflects the behavior of the individual's tendency to choose the destination whose benefit is higher than the benefits of the origin and the intervening opportunities. Parameter $\beta$ reflects the behavior of the individual's tendency to choose the destination whose benefit is higher than  the benefit of the origin, and the benefit of the origin is higher than the benefits of the intervening opportunities. According to the above assumption, the probability that location $j$ is selected by the individual at location $i$ is
\begin{equation}
\label{eq1}
Q_{ij}=\int_0^{\infty}{\mathrm{Pr}_{m_{i}+{\alpha}{\cdot}s_{ij}}(z)}
{\mathrm{Pr}_{{\beta}{\cdot}s_{ij}}(<z)}\mathrm{Pr}_{m_{j}}(>z) \mathrm{d}z,
\end{equation}
where $m_{i}$ is the number of opportunities at location $i$, $m_{j}$ is the number of opportunities at location $j$, $s_{ij}$ is the number of intervening opportunities \cite{S40} (i.e., the sum of the number of opportunities at all locations whose distances from $i$ are shorter than the distance from $i$ to $j$), ${\mathrm{Pr}_{m_{i}+{\alpha}{\cdot}s_{ij}}(z)}$ is the probability that the maximum benefit obtained after $m_{i}+{\alpha}{\cdot}s_{ij}$ samplings is exactly $z$, ${\mathrm{Pr}_{{\beta}{\cdot}s_{ij}}(<z)}$ is the probability that the maximum benefit obtained after ${\beta}{\cdot}s_{ij}$ samplings is less than $z$, ${Pr}_{m_{j}}(>z)$ is the probability that the maximum benefit obtained after $m_{j}$ samplings is greater than $z$, $\alpha$ and $\beta$ are both non-negative and $\alpha+\beta\leq1$. 

{\color{black}Since $\mathrm{Pr}_{x}(<z)=p(<z)^{x}$, we obtain 
\begin{equation}
\label{eq2}
\mathrm{Pr}_{x}(z)=\frac{\mathrm{d}\mathrm{Pr}_{x}(<z)}{\mathrm{d}z}=x p(<z)^{x-1}\frac{\mathrm{d}p(<z)}{\mathrm{d}z}.
\end{equation}
Eq. (\ref{eq1}) can be written as 
\begin{equation}
\label{eq3}
\begin{aligned}
Q_{ij}=&\int_0^{\infty}{\mathrm{Pr}_{m_{i}+{\alpha}{\cdot}s_{ij}}(z)}
{\mathrm{Pr}_{{\beta}{\cdot}s_{ij}}(<z)}\mathrm{Pr}_{m_{j}}(>z) \mathrm{d}z\\
=&(m_i+{\alpha} s_{ij}) \int_0^{1}(p(<z)^{m_i+({\alpha}+{\beta})s_{ij}-1}\\
&-p(<z)^{m_i+({\alpha}+{\beta})s_{ij}+m_{j}-1})\mathrm{d}p(<z)\\
=&(m_i+{\alpha} s_{ij}) (\frac{p(<z)^{m_i+({\alpha}+{\beta})s_{ij}}}{m_i+({\alpha}+{\beta})s_{ij}}\Big\vert_0^{1}\\
&-\frac{p(<z)^{m_i+({\alpha}+{\beta})s_{ij}+m_j}}{m_i+({\alpha}+{\beta})s_{ij}+m_j}\Big\vert_0^{1})\\
=&\frac{({m_i+{\alpha}s_{ij}}){m_j}}{[{m_i+({\alpha}+{\beta})s_{ij}}][{m_i+({\alpha}+{\beta})s_{ij}+m_j}]}.
\end{aligned}
\end{equation}}
Then, the probability of the individual at location $i$ choosing location $j$ is
\begin{equation}
\label{eq4}
P_{ij}=\frac{Q_{ij}}{\sum\limits_j Q_{ij}}\propto\frac{({m_i+{\alpha}s_{ij}}){m_j}}{[{m_i+({\alpha}+{\beta})s_{ij}}][{m_i+({\alpha}+{\beta})s_{ij}+m_j}]}.
\end{equation}
Further, if we know the total number of individuals $O_i$ who travel from location $i$, the flux $T_{ij}$ from location $i$ to location $j$ can be calculated as
\begin{equation}
\label{eq5}
T_{ij}=O_iP_{ij}=
O_i\frac{\frac{({m_i+{\alpha}s_{ij}}){m_j}}{([{m_i+({\alpha}+{\beta})s_{ij}}][{m_i+({\alpha}+{\beta})s_{ij}+m_j}])}}{\sum\limits_j\frac{({m_i+{\alpha}s_{ij}}){m_j}}{([{m_i+({\alpha}+{\beta})s_{ij}}][{m_i+({\alpha}+{\beta})s_{ij}+m_j}])}}.
\end{equation}
This is the final form of the model and we name it the universal opportunity (UO) model.

\begin{figure} 
\flushright
\includegraphics[width=1.02\linewidth]{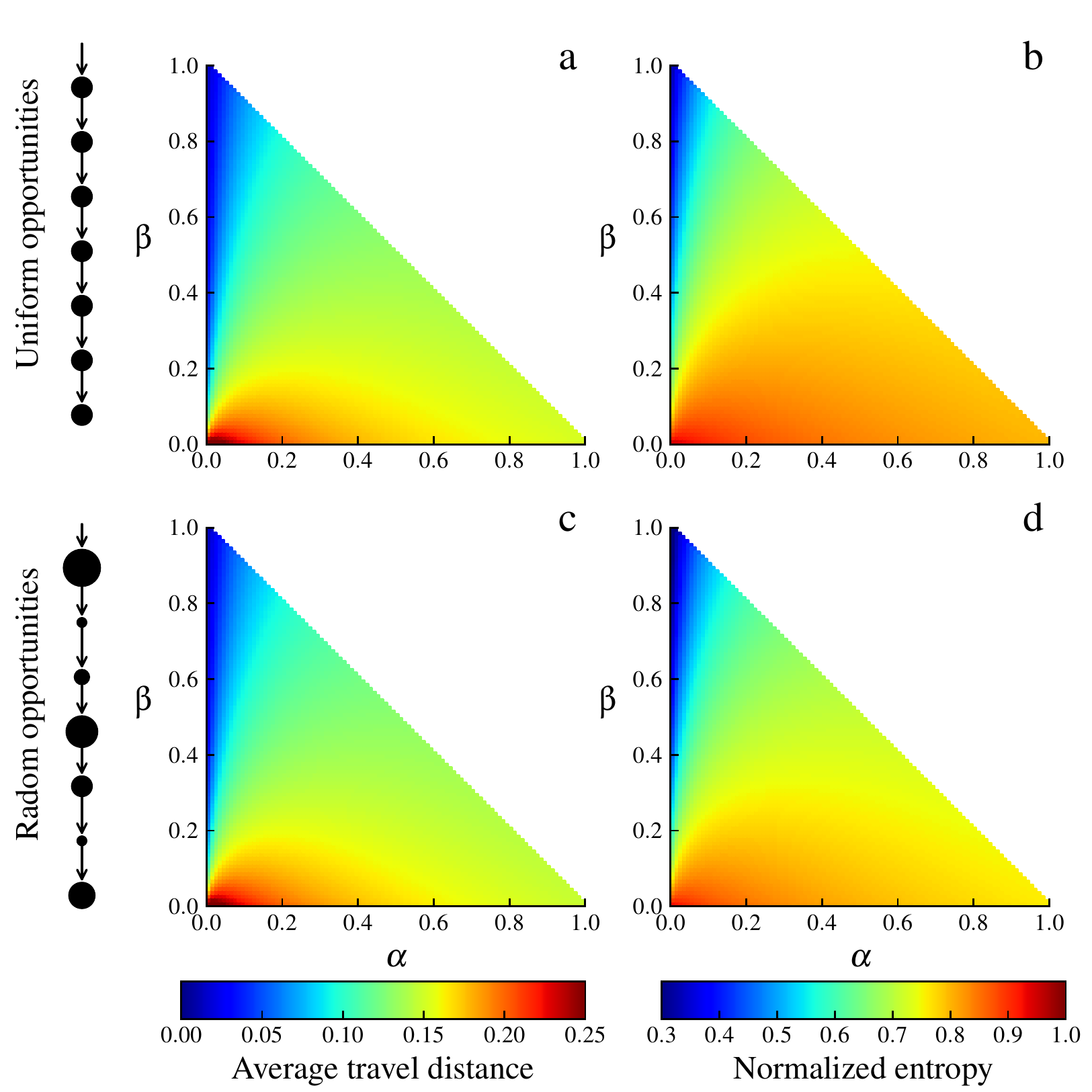}
\caption{{\bf Average travel distance and normalized entropy versus different parameter combinations.} ({\it a}-{\it b}) Average travel distance and normalized entropy values corresponding to different parameter combinations. Here, the number of destination opportunities is a uniform distribution. ({\it c}-{\it d}) Same average travel distance and normalized entropy values as in ({\it a}-{\it b}), but the number of destination opportunities is a random distribution.}
\label{fig1}
\end{figure}

{\color{black} The $\alpha$ and $\beta$ parameters in the UO model reflect the two behavioral tendencies of the individual when choosing potential destinations (where the opportunity benefit is higher than the benefit of the origin). From Eq. (\ref{eq3}), we can see that the larger the value of parameter $\alpha$, the greater the probability that distant potential destinations will be selected by the individual. We name this behavioral tendency the exploratory tendency. On the other hand, the larger the value of parameter $\beta$, the greater the probability that near potential destinations will be selected by the individual. We name this behavioral tendency the cautious tendency.}	
We choose average travel distance and normalized entropy as two fundamental metrics to discuss the influence of two parameters $\alpha$ and $\beta$ on individual destination selection behavior. The average travel distance reflects the bulk density of  individual destination selection \cite{BHG06,GHB08,Y13,roth11}, and normalized entropy reflects the heterogeneity of individual destination selection \cite{ea10}. 
As shown in Fig. \ref{fig1}, the two fundamental metrics have the same  regularities with a change in two parameters, whether the number of destination opportunities is a uniform or random distribution. When $\alpha=0$, $\beta=1$, the average travel distance is the shortest, and the normalized entropy value is the smallest; when $\alpha=0$, $\beta=0$, the average travel distance is the longest, and the normalized entropy value is the largest. From the definitions of the two parameters, we can easily explain the reasons for the regularities. When $\alpha$ is closer to 0, $\beta$ is closer to 1, the individual {\color{black} is more cautious, and the probability of choosing near potential destinations is higher}, so the shorter the average travel distance and the stronger the heterogeneity. {\color{black} When $\alpha$ is closer to 1, $\beta$ is closer to 0, the individual is more exploratory, and the probability of choosing distant potential destinations is higher, so the average distance is increased while the heterogeneity is decreased.} When $\alpha$ and $\beta$ are both closer to 0, the individual attaches more importance to the benefit that the location brings to him/her and does not care about the order of locations, so the longer the average travel distance and the stronger the homogeneity.

Moreover,  when $\alpha$ and $\beta$ take extreme values (i.e., the three vertices of the triangle in Fig. \ref{fig1}), we can derive three special human mobility models. When $\alpha=0$, $\beta=0$, we name this model the opportunity only (OO) model. In this model, the individual chooses the location whose benefit is higher than the benefit of the origin. Then, the probability of the individual at location $i$ choosing location $j$ as the destination is
\begin{equation}
\label{eq6}
P_{ij}=\frac{m_j/(m_i+m_j)}{\sum\limits_j m_j/(m_i+m_j)}\propto\frac{m_j}{{m_i+m_j}}.
\end{equation}
When $\alpha=1$, $\beta=0$, our model can be simplified to the OPS model, in which the individual chooses the location whose benefit is higher than the benefit of the origin and the benefits of the intervening opportunities. Then, the probability of the individual at location $i$ choosing location $j$ as the destination is
\begin{equation}
\label{eq7}
P_{ij}=\frac{m_j/(m_i+s_{ij}+m_j)}{\sum\limits_j m_j/(m_i+s_{ij}+m_j)}\propto\frac{m_j}{{m_i+s_{ij}+m_j}}.
\end{equation}
When $\alpha=0$, $\beta=1$, our model can be simplified to the radiation model, in which the individual chooses the location whose benefit is higher than the benefit of the origin and the benefits of the intervening opportunities are lower than the benefit of the origin. Then, the probability of the individual at location $i$ choosing location $j$ as the destination is
\begin{equation}
\label{eq8}
\begin{aligned}
P_{ij}=&\frac{m_im_j/[({m_i+s_{ij}})({m_i+s_{ij}+m_j})]}{\sum\limits_j m_im_j/[({m_i+s_{ij}})({m_i+s_{ij}+m_j})]}\\
\propto&\frac{m_im_j}{({m_i+s_{ij}})({m_i+s_{ij}+m_j})}.
\end{aligned}
\end{equation}
From equations (\ref{eq6})-(\ref{eq8}), we can see that the OO model, the OPS model and the radiation model are all special cases of our UO model.

\subsection{Prediction}
\begin{figure*} 
\centering
\includegraphics[width=1.0\linewidth]{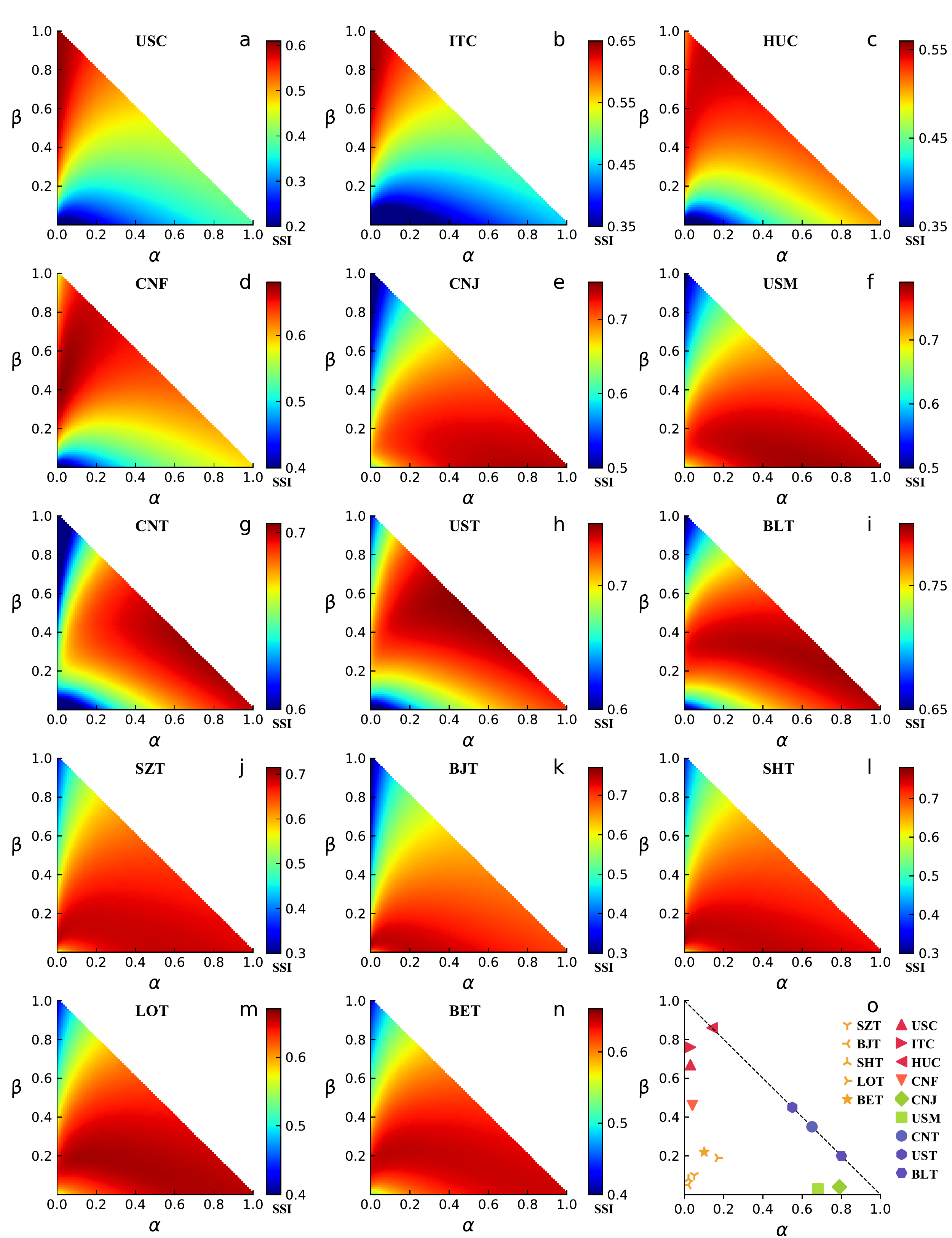}
\caption{{\bf  Results for empirical data sets.} ({\it a}-{\it {\color{black} n})}  We exploit SSI to calculate the similarity between the real mobility matrix and the predicted mobility matrix under different parameter combinations for the {\color{black}fourteen} data sets. Here, the color bar represents the SSI, where a dark red (blue) dot indicates a higher (lower) SSI.  ({\it {\color{black} o}})  The optimal values of the parameters $\alpha$ and $\beta$ correspond to the highest SSI for the {\color{black}fourteen} data sets.}
\label{fig2}
\end{figure*}

We use {\color{black} fourteen} empirical data sets, including commuting trips between United States' counties (USC),  {\color{black} commuting trips between the provinces of Italy (ITC), commuting trips between the subregions of Hungary(HUC),} freight between Chinese cities (CNF), internal job hunting in China (CNJ), internal migrations in the US (USM), intercity travels in China (CNT), intercity travels in the US (UST),  {\color{black} intercity travels in Belgium (BLT),} intracity trips in Suzhou (SZT),  {\color{black} intracity trips in Beijing(BJT), intracity trips in Shenzhen (SHT), intracity trips in London (LOT) and intracity trips in Berlin (BET)} (see  {\bf Methods}), to validate the predictive ability of the UO model.
We first extract the flux $T_{ij}$  from location $i$ to location $j$ from the data set and obtain the real mobility matrix. Then, we exploit the S{\o}rensen similarity index \cite{YZFDW14} (SSI, see {\bf Methods}) to calculate the similarity between the real mobility matrix and the mobility matrix predicted by the UO model under different parameter combinations. The results are shown in Fig. \ref{fig2}. {\color{black}Figure \ref{fig2}{\it  o } shows the optimal values of the parameter $\alpha$ and $\beta$ corresponding to the highest SSI for the fourteen data sets. }

\begin{figure*} 
\flushleft
\includegraphics[width=0.96\linewidth]{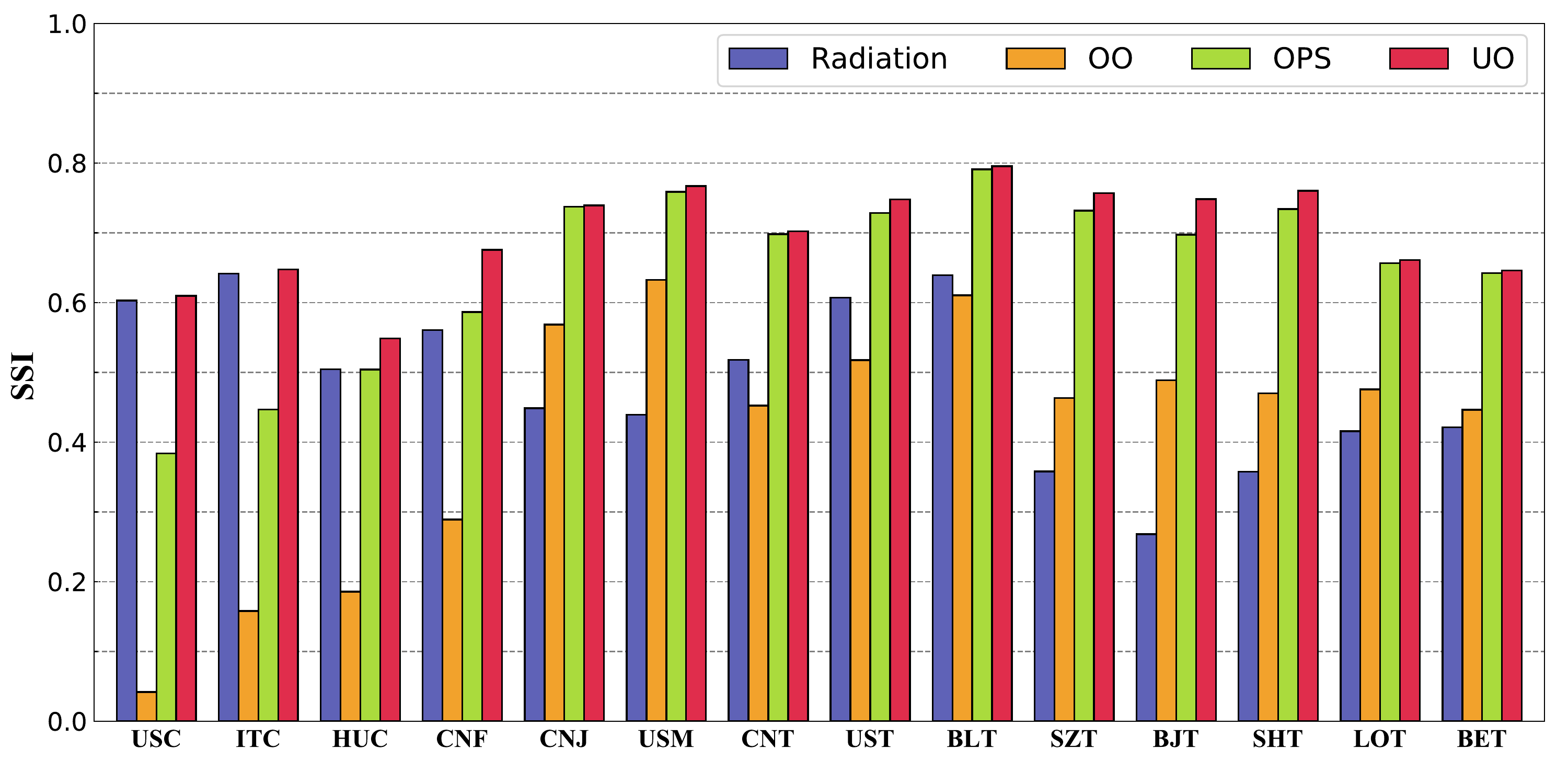}
\caption{{\bf Comparing predicting accuracy of the UO model, the radiation model, the OPS model and the OO model  in terms of SSI.}}
\label{fig-3}
\end{figure*}

\begin{table*}\centering

	\caption{{\color{black}{\bf Comparison of models prediction accuracy.} SSI is the S{\o}rensen similarity index between the real mobility matrix and the mobility matrix predicted by different models. RMSE is the root mean square error of predicted mobility matrix. UO, RM, OPS, and OO stand for the universal opportunity model, the radiation model, the opportunity priority selection model and  the opportunity only model, respectively.} }
	\label{tb-1}
    	{\color{black}
	\begin{tabular}{p{1.6cm}p{1.6cm}p{1.6cm}p{1.6cm}p{1.6cm}p{1.99cm}
			p{1.99cm}p{1.99cm}p{1.99cm}}
		\hline
		Data set & SSI-UO &  SSI-RM  & SSI-OPS  &  SSI-OO 
		& RMSE-UO & RMSE-RM & RMSE-OPS & RMSE-OO \\ 
		\hline
		USC & 0.610 &  0.603  & 0.384  &  0.042 	
		& 2158.766 & 2308.054 & 2948.205 & 3654.402 \\ 
		ITC & 0.648 &  0.641  & 0.447  &  0.158
		& 1600.862 & 1696.488 & 2132.627 & 3033.019 \\
		HUC & 0.549 &  0.504  & 0.504  &  0.186   
		& 477.254 & 546.878 & 429.377 & 612.904 \\
		CNF & 0.676 &  0.561  & 0.587  &  0.289   
		& 111.201 & 184.724 & 128.789 & 183.655 \\
		CNJ & 0.739 &  0.449  & 0.738  &  0.567   
		& 185.709 & 481.072 & 189.816 & 297.379 \\
		USM & 0.767 &  0.434  & 0.759  &  0.632   
		& 1126.110 & 3275.661 & 1218.255 & 1521.585 \\
		CNT & 0.702 &  0.518  & 0.698  &  0.452  
		& 441.063 & 829.869 & 438.463 & 731.153 \\
		UST & 0.748 &  0.607  & 0.729  &  0.518  
		& 55.851 & 95.013 & 65.513 & 115.795 \\
		BLT & 0.796 &  0.639  & 0.791  &  0.611   
		& 26.236 & 58.641 & 26.339 & 48.080 \\
		SZT & 0.757 &  0.358  & 0.732  &  0.463   
		& 7.871 & 47.801 & 9.133 & 12.553 \\
		BJT & 0.748 &  0.268  & 0.697  &  0.489   
		& 6.567 & 68.039 & 12.291 & 12.040 \\
		SHT & 0.760 &  0.358  & 0.734  &  0.470   
		& 48.196 & 368.901 & 71.152 & 91.000 \\
		LOT & 0.661 &  0.416  & 0.657  &  0.476   
		& 4.309 & 20.031 & 4.603 & 8.104 \\
		BET & 0.646 &  0.421  & 0.642  &  0.447   
		& 3.288 & 11.271 & 3.356 & 5.323 \\
		\hline
	\end{tabular}
}
\end{table*}

{\color{black} It can be seen from Fig. \ref{fig2}{\it a}-{\it d} that for USC, ITC, HUC and CNF, when $\alpha$ is close to 0 and $\beta$ is close to 1, the SSI is relatively large. The reason is that for commuting data sets (USC, ITC and HUC), the commuting distance or time is very important for commuters. As a result, most people tend to choose near potential destinations when finding a job based on their place of residence or adjusting their place of residence after finding a job. This cautious destination selection tendency also exists in freight. Freight to far destinations will lead to an increase in transportation costs and a decrease in the freight frequency, which will have a negative impact on freight revenue. Thus, unless the destination opportunity benefit is very high, the individual tends to choose a near destination rather than a far destination for freight. For the migration and job hunting data sets (USM and CNJ), when $\alpha$ is close to 1 and $\beta$ is close to 0, the SSI is relatively large, as shown in Fig. \ref{fig2}{\it e}-{\it f}. The reason is that both job seekers and migrants pay more attention to the destination opportunity benefit rather than the distance to the destination. In other words, they are more exploratory but less cautious. Even if a high benefit destination is far away, it will still be selected by individuals with a relatively high probability. The reason is that the distance to the destination has a smaller impact on long temporal scale mobility behaviors, such as migration and job hunting, than on daily commuting behaviors. For intercity travel  data sets (CNT, UST and BLT), when $\alpha$ and $\beta$ are both near the middle of the diagonal line of the triangle, the SSI is relatively large, as shown in Fig.  \ref{fig2}{\it g}-{\it i}. For most people, intercity travel is occasional and not as frequent as commuting. Travelers are less inclined than commuters to choose near potential destinations but they tend to explore distant potential destinations. Thus, the exploratory tendency parameter $\alpha$ of intercity travels is much larger than that of commuting. On the other hand, the importance of the travel cost of intercity travels is higher than that of the cost of migration. Thus, the cautious tendency parameter $\beta$ of intercity travels is larger than that of migration. For intracity trips data sets (SZT, BJT, SHT, LOT and BET), when $\alpha$ and $\beta$ are both close to 0, the SSI is relatively large, as shown in Fig. \ref{fig2}{\it j}-{\it n}.  The reason is that  compared with the intercity mobility behavior on a large spatial scale, the spatial scale of intracity mobility behavior is small. In this scenario, the individual is not necessarily concerned about the travel distance and focuses more on the benefit that the location will directly bring to him/her. Thus, the optimal values of $\alpha$ and $\beta$ are both close to 0, as shown in Fig. \ref{fig2}{\it o}.
}

We next compare the predictive accuracy of the mobility fluxes of the UO model with the radiation model, the OPS model and the OO model. In terms of SSI, as
shown in Fig. \ref{fig-3} {and \color{black} Table \ref{tb-1}}, the UO model performs best. However,  {\color{black} the radiation model and the OPS model can provide only relatively accurate predictions for some data sets.  For example, the radiation model can predict commuting and freight trips relatively accurately but cannot accurately predict other types of mobility. The reason is that the individual tends to choose near potential destinations rather than distant potential destinations in commuting and freight, where travel costs are more important. From Fig. \ref{fig2}o, we can see that for commuting and freight data sets, the optimal parameter $\beta$ (which reflects the individual's cautious tendency) of the UO model is close to 1, and the  optimal parameter $\alpha$ (which reflects the individual's exploratory tendency) is close to 0. Therefore, the prediction accuracy of the radiation model in which the individual only chooses the closest potential destination (i.e., $\alpha=0, \beta=1$) is close to that of the UO model in commuting and freight data sets. However, the prediction accuracy of the radiation model is considerably lower than that of the UO model in job hunting, migration and noncommuting travel data sets. The reason is that the individual is more likely to choose distant potential destinations in these data sets. In these cases, the prediction accuracy of the OPS model, in which the individual tends to choose distant potential destinations, is closer to that of the UO model.} 
{\color{black} We further use a frequently used  statistical index, named the root mean square error (RMSE), to measure the prediction errors of the UO model and the other three models, and Table \ref{tb-1} lists the results . From the table, we can see that in most cases, the RMSE of the UO model is smaller than that of the other benchmark models, although the RMSE is not the parameter optimization objective of the UO model.} 
These results prove that the three models only capture the individual's destination selection behavior at a specific spatiotemporal scale. Yet our UO model can accurately describe the individual's destination selection behavior at different spatiotemporal scales.

\section{Discussion}
	
Although previous IO class models are widely used to predict the mobility of people between locations \cite{SGMB12,SMN13,KLGQ15,VTN18,var16,REWGT14,YZFDW14,YWGL17,sim15,LY19}, these models can only achieve accurate prediction at specific spatiotemporal scales. In this paper, we developed a UO model to predict human mobility at different spatiotemporal scales.
Our model establishes a new framework in IO class models and covers the classical radiation model \cite{SGMB12} and the OPS model \cite{LY19}. {\color{black} Although the UO model has two parameters, they are different from the parameters in some regression analysis models or  machine learning models in the sense that they simply  improve the prediction accuracy of the model. These two parameters essentially describe the two tendencies, i.e., exploratory tendency and cautious tendency, of an individual's destination selection behavior. They not only enable the UO model to better predict human mobility at different spatiotemporal scales than the parameter-free models but also help us better understand the underlying mechanism of the individual's destination selection behavior in different types of human mobility.}

Many phenomena in complex system field are strongly related to human mobility \cite{BBG17}.
For example,  the spread of disease is directly affected by human travel distance between locations and the population size of locations  \cite{hu004,eu004,bal09,watt05,kra16,kit10,vib06}. The UO model can accurately describe the individual's destination selection behavior at different spatiotemporal scales, which has potential applications for  understanding the spread of disease within humans.
Not only that, but the IO model can also describe an individual's selection behavior in social networks such as friend networks and scientific collaboration networks.
In friend networks,  the individual tends to choose friends who are close to him/her and  have a high sense of identity \cite{sim15,ill13}. In scientific collaboration networks, the individual tends to choose nearby scholars who have high scientific influence \cite{pan12}.
These phenomena indicate that when one seeks to build beneficial ties, she/he will take into account both the distance and the benefits of the opportunities. The UO model can describe the individual's interactive object selection behavior, providing a new perspective for social network analysis.

Despite its fine performance in predicting human mobility, the UO model has room for further improvements. For example,  most existing IO class models use an agent to represent all of the individuals and neglect the diversity of individual selection behavior \cite{Y13,song10,bj11,panl15,Gal16,zhao16}. 
Building mobility prediction model for each individual may reflect the diversity in detail.
However, it is extremely cumbersome and cannot grasp the commonality among individuals' mobility patterns.
One possible approach is first clustering  individuals according to their mobility behavior characteristics \cite{lou15,lian11,lian14}, then expanding our UO model for different classes of individuals, which may  more accurately predict human mobility.

\section{Material and methods}
\subsection{Data sets}

(1) {\color{black}Commuting trips. The commuting trips data sets include the commuting trips between United States' counties \cite{SGMB12} (USC) , the commuting trips between the provinces of Italy \cite{VTN18} (ITC) and the commuting trips between the subregions of Hungary \cite{VTN18} (HUC), which were downloaded from http://www.census.gov/population/www/cen2000/\\com-muting/index.html, http://www.stat.it/storage/\\cartografia/matrici\underline{\hbox to 0.15cm{}} pendolarismo/matrici\underline{\hbox to0.15cm{}}pendolarismo\underline{\hbox to 0.15cm{}} 2011.zip and http://www.ksh.hu, respectively. Since we focus on mobility among zones(counties, provinces or subregions), all the residences/workplaces within a zone are regarded as the same with an identical zone label. Then, we can accumulate the total number $T_{ij}$ of trips from zone   $i$ to zone $j$,  which is also carried out in the  following data sets.}

(2) Freight between Chinese cities (CNF). The CNF data set is extracted from the travel records of freight between Chinese cities from 19 May 2015 to 23 May 2015. When freight is loaded or unloaded, the coordinates and time are recorded automatically by a GPS-based device installed in the truck. All the loading/unloading locations within a city are regarded as the same with an identical zone label.

(3) Internal job hunting in China (CNJ). The CNJ data set is extracted from more than 160 million job hunters' resumes from 2006 to 2016 and was downloaded from https://www.zhaopin.com. The resumes contain job hunter work experience, from which we can obtain a job hunter's former workplaces. All the workplaces within a city are regarded as the same with an identical zone label. 

(4) Internal migrations in the US (USM). The USM data set is extracted from the Statistics of Income Division of the Internal Revenue Service (IRS) in the US from 2011 to 2012 and was downloaded from https://www.irs.gov/statistics/soi-tax-stats-migration-data. The IRS contains records of all individual income tax forms filed in each year, from which we can determine who has or has not, moved residence/workplace locations in the intervening fiscal year \cite{BBG17}. All the residence/workplace locations within a state are regarded as the same with an identical zone label.

(5) {\color{black}Intercity travels. The intercity travels data sets include intercity travels in China (CNT), intercity travels in the US (UST) and intercity travels in Belgium (BLT).} The CNT data set is extracted from  check-in records of the Sina Weibo website for users in mainland China \cite{YWGL17}. The UST data set is extracted from check-in records of the  Foursquare website for users in the continental US \cite{lev12}. {\color{black}The BLT data set is extracted from check-in records of the website Gowalla for users in Belgium \cite{cho11}.}
{\color{black}These data sets contain} each user's spatial and temporal information, from which we can obtain the user's location. All the check-in locations within a city are regarded as the same with an identical zone label.

(6) {\color{black}Intracity trips. The intracity trips data sets include 
intracity trips in Suzhou (SZT), intracity trips in Beijing (BJT), intracity trips in Shenzhen (SHT), intracity trips in London (LOT) and intracity trips in Berlin (BET).}
The SZT data set is extracted from the mobile phone call detail records in Suzhou, a city of China. The data contains the time and positions of users making phone calls or sending text messages. {\color{black}The BJT data set \cite{liang12} and the SHT data set \cite{liang12} are extracted from the travel records of taxi passengers in Beijing and Shenzhen, respectively. When a passenger gets on or gets off a taxi, the coordinates and time are recorded automatically by a GPS-based device installed in the taxi.
The LOT data set \cite{cho11} and the BET data \cite{cho11} set are extracted from checkin records at Gowalla in London and Berlin. Because of the absence of natural partitions in cities (in contrast to states or counties), the city is divided  into zones, each of which is 1 km $\times$ 1 km (for SZT is 0.01 longitude $\times$ 0.01 latitude).}
All the locations within a zone are regarded as the same with an identical zone label \cite{YZFDW14}.

\subsection{Normalized entropy}
We use normalized entropy to reflect the heterogeneity of individual destination selection
\begin{equation}
\label{eq9}
E_{i}=\dfrac{-\sum\limits_{j=1}^{N}p_{ij}\log(p_{ij})}{\log(p_{ij})},
\end{equation}
where $E_{i}$ is the normalized entropy of location $i$, $p_{ij}$ is the probability that the individual at location $i$ chooses location $j$ as his/her destination, and $N$ is the number of locations.
\subsection{S{\o}rensen similarity index}
The S{\o}rensen similarity index \cite{S48} is a similarity measure between two samples. Here, we apply a modified version \cite{YZFDW14} of the index to measure whether real fluxes are correctly reproduced (on average) by theoretical models, defined as
\begin{equation}
\label{eq10}
\mathrm{SSI} = \frac{1}{N(N-1)}\sum^{N}_{i}{\sum^{N}_{j \neq i}{\frac{2 \min (T_{ij},T^{'}_{ij})}{T_{ij}+T^{'}_{ij}} }},
\end{equation}
where $N$ is the number of locations, $T_{ij}$ is the predicted flux from location $i$ to $j$ and $T^{'}_{ij}$ is the empirical flux. Obviously, if each $T_{ij}$ is equal to $T^{'}_{ij}$ the index is 1, and if all $T_{ij}$ are far  from the real values, the index is close to 0.

\noindent\textbf{Acknowledgements:} 
X.-Y.Y. was supported by NSFC under grant nos. 71822102, 71621001 and 71671015.

\end{document}